\documentclass[11pt]{revtex4-1}

\usepackage{graphicx}

\begin{document}

\title{A REDEFINITION OF HAWKING TEMPERATURE ON THE EVENT HORIZON: THERMODYNAMICAL EQUILIBRIUM}

\author{Subhajit Saha\footnote {subhajit1729@gmail.com}}
\author{Subenoy Chakraborty\footnote {schakraborty.math@gmail.com}}

\affiliation{Department of Mathematics, Jadavpur University, Kolkata-700032, West Bengal, India.}

%%%%%%%%%%%%%%%%%%%%%%%%%%%%%%%%%%%%%%%%%%%%%%%%%%%%%%%%%%%%%%%%%%%%%%%%%%%%%%%%%%%%%%%%%%%%%%%%%%%%%%%%%%%%%%%%%%

\begin{abstract}

In this article we have used the recently introduced redefined Hawking temperature on the event horizon and investigated whether the generalised second law of thermodynamics (GSLT) and thermodynamic equilibrium holds for both the event and the apparent horizons. Here we have considered FRW universe and examined the GSLT and thermodynamic equilibrium with three examples. Finally, we have concluded that from the thermodynamic viewpoint, the universe bounded by the event horizon is more realistic than that by the apparent horizon at least for some examples.\\\\
Keywords: Thermodynamical equilibrium, Hawking temperature, Bekenstein system.\\
PACS Numbers: 98.80.Cq, 98.80.-k

\end{abstract}

\maketitle

%%%%%%%%%%%%%%%%%%%%%%%%%%%%%%%%%%%%%%%%%%%%%%%%%%%%%%%%%%%%%%%%%%%%%%%%%%%%%%%%%%%%%%%%%%%%%%%%%%%%%%%%%%%%%%%%%%%%%%%%%%%%
%\myclassification{98.80.Cq, 98.80.-k}\\\\
%%%%%%%%%%%%%%%%%%%%%%%%%%%%%%%%%%%%%%%%%%%%%%%%%%%%%%%%%%%%%%%%%%%%%%%%%%%%%%%%%%%%%%%%%%%%%%%%%%%%%%%%%%%%%%%%%%%%%%%%%%%%

\section{INTRODUCTION}

A detailed comparative study of the apparent and event horizon for the homogeneous and isotropic model of the universe in accelerating phase has been done by Wang {\it et al.} [1] particularly the validity of the thermodynamical laws at these horizons. They showed the validity of both the first and second law of thermodynamics on the dynamical apparent horizon but were unsuccessful at the event horizon. Also they were not able to restore the thermodynamical laws by redefining any parameter involved. As a result they concluded that cosmological event horizon is unphysical from thermodynamical point of view. Further, they had shown that although universe bounded by the apparent horizon is a Bekenstein system but due to breakdown of the Bekenstein bounds universe bounded by the event horizon is no longer a Bekenstein system.

In recent past, we have shown [2] the validity of the second law of thermodynamics on the event horizon assuming the validity of the first law with some restrictions and that is true in any gravity theory. Very recently [3], we are able to deduce the first law of thermodynamics on the event horizon for simple DE fluid by appropriately defining the Hawking temperature on the event horizon. In that work, we start with FRW model of the universe and the metric can locally be expressed in the form
\begin{equation}
ds^{2}=h_{i j}\left(x^{i}\right)dx^{i} dx^{j}+R^{2} d\Omega_2^{2}
\end{equation}
where $i$, $j$ can take values $0$ and $1$.\\
The two dimensional metric
\begin{equation}
d\gamma^{2}=h_{i j}\left(x^{i}\right)dx^{i} dx^{j}
\end{equation}
where
\begin{equation}
h_{i j}=diag\left\{-1, \frac{a^2}{1-\kappa r^2}\right\}
\end{equation}
is referred to as the normal metric with $x^{i}$ being the associated co-ordinates ($x^0=t,x^1=r$). $R=ar$ is the area radius, considered as a scalar field in the normal two-dimensional space. One can define another scalar associated with this normal space as
\begin{equation}
\chi(x)=h^{i j}\left(x^{i}\right)\partial_{i}R \partial_{j}R=1-\left(H^2+\frac{\kappa}{a^2}\right)R^2
\end{equation}
with $\kappa=0,+1,-1$ for flat, closed and open model respectively.\\
The apparent horizon, a null surface is defined at the vanishing of this scalar, {\it i.e.,} $\chi(x)=0$ and we have
\begin{equation}
R_A=\frac{1}{\sqrt{H^2+\frac{\kappa}{a^2}}}.
\end{equation}
Consequently, the surface gravity on the horizon is defined as 
\begin{equation}
\kappa_{A}=-\frac{1}{2}\frac{\partial \chi}{\partial R}|_{R=R_{A}}=\frac{1}{R_A}
\end{equation}
and hence we get the usual Hawking temperature on the apparent horizon as
\begin{equation}
T_A=\frac{|\kappa_A|}{2\pi}=\frac{1}{2\pi R_A}
\end{equation}
Now instead of defining the Hawking temperature on the event horizon as $T_E=\frac{1}{2\pi R_E}$ in analogy with the usual Hawking temperature $T_A=\frac{1}{2\pi R_A}$ on the apparent horizon, we start from the definition as $T_h=\frac{|\kappa_h|}{2\pi}$ with $\kappa_h=-\frac{1}{2}\frac{\partial \chi}{\partial R}|_{R=R_h}$. As a result we have [3]
\begin{equation}
\kappa_E=\frac{R_E}{R_A^2}
\end{equation}
and
\begin{equation}
T_E=\frac{|\kappa_E|}{2\pi}=\frac{R_E}{2\pi R_A^2}.
\end{equation}
In the context of the {\it Wilkinson Microwave Anistropy Probe data} [4] we choose flat FRW model for which the two horizons are connected by the inequality
\begin{equation}
R_A=\frac{1}{H}=R_H<R_E
\end{equation}
and the Hawking temperature on the horizons simplifies to
\begin{equation}
T_E=\frac{H^2R_E}{2\pi}>\frac{H}{2\pi}=T_A
\end{equation}

%%%%%%%%%%%%%%%%%%%%%%%%%%%%%%%%%%%%%%%%%%%%%%%%%%%%%%%%%%%%%%%%%%%%%%%%%%%%%%%%%%%%%%%%%%%%%%%%%%%%%%%%%%%%%%%%%%

\section{VALIDITY OF GSLT AND THERMODYNAMIC EQUILIBRIUM}

In the present work, using this modified form of the Hawking temperature on the event horizon we examine the validity of the second law of thermodynamics. The main motivation is to investigate whether the physical system namely universe bounded by the horizon (event or apparent) approaches an equilibrium configuration or not. According to thermodynamics, the equilibrium configuration of an isolated macroscopic physical system corresponds to the maximum entropy state(consistent with the constraints imposed on the system). As a result, the entropy function cannot decrease, {\it i.e.,}~~$\dot{S}\geq0$. Also $\ddot{S}<0$ holds due to the fact that the entropy function attains a maximum [5]. Earlier Pavon and Zimdahl [5] have studied thermodynamical equilibrium for the universe bounded by the apparent horizon when it approaches to future singularity by some examples. Here we deal with both the horizons in a general way and consider universe filled with exotic matter ({\it i.e.,} dark fluids). So for the generalized second law of thermodynamics as well as for thermodynamic equilibrium the entropy functions should satisfy 
\begin{equation}
i)~~\dot{S}_h+\dot{S}_{fh}\geq0~~~,~~~ii)~~\ddot{S}_h+\ddot{S}_{fh}<0
\end{equation}
where $S_h$ and $S_{fh}$ are the entropies of the horizon and that of the fluid within it respectively.\\
To determine the entropy on the apparent horizon we start with Bekenstein's entropy-area relation on the apparent horizon as
\begin{equation}
S_A=\frac{A_A}{4G}=\frac{\pi R_A^2}{G},
\end{equation}
so 
\begin{equation}
\dot{S}_A=\frac{2\pi R_A\dot{R}_{A}}{G}~~~,~~~\ddot{S}_A=\frac{2\pi}{G}(\dot{R}_A+R_A\ddot{R}_A).
\end{equation}
But according to Wang {\it et al.} [1] the system bounded by the event horizon is not a Bekenstein system. So it is not reasonable to use Bekenstein's entropy-area relation on the event horizon. Thus to determine entropy on the event horizon we start with the first law of thermodynamics ({\it i.e.,} Clausius relation) on it as 
\begin{equation}
T_EdS_E=\delta Q=-dE=4\pi {R_E}^3(\rho +p)Hdt.
\end{equation}
Using equation (11) for $T_E$ we have
\begin{equation}
\dot{S}_E=\frac{8\pi ^2{R_E}^2(\rho +p)}{H}
\end{equation}
and
\begin{equation}
\ddot{S}_E=8\pi ^2(\rho +p)R_E\lbrace{-3(1+\frac{\dot{p}}{\dot{\rho}})+(1+\frac{p}{\rho})+2(R_E-\frac{1}{H})\rbrace}.
\end{equation}
To determine the entropy variation of the fluid bounded by the horizon we start with the Gibb's relation [1,6]
\begin{equation}
T_{fh}dS_{fh}=dE+pdV
\end{equation}
where $V=\frac{4}{3}\pi R_h^3$ is the volume of the fluid, $E=\rho V$ is the total energy, $\rho$, $p$ are the usual energy density and thermodynamic pressure of the fluid and $T_{fh}$ is the temperature of the fluid bounded by the horizon. Hence from (18) the explicit form of the variation of the entropy of the fluid is given by
\begin{equation}
\dot{S}_{fh}=\frac{4\pi R_h^2(\rho +p)}{T_{fh}}(\dot{R}_h-HR_h)
\end{equation}
Again differentiating once we obtain
\begin{equation}
\ddot{S}_{fh}=\frac{4\pi R_h(\rho +p)}{T_{fh}}[(2\dot{R}_h-3HR_h)(\dot{R}_h-HR_h)+R_h\lbrace{\ddot{R}_h}-(R_h\dot{H}+\dot{R}_hH)\rbrace],
\end{equation}
where we have used the energy conservation relation
\begin{equation}
\dot{\rho}+3H(\rho +p)=0.
\end{equation}
In the following we choose $T_{fh}=T_h$, {\it i.e.,} temperature of the fluid same as temperature of the horizon for equilibrium configuration.

We shall now consider the entropy variations for the two horizons seperately.

\subsection{Apparent horizon:}

\begin{equation}
R_A=\frac{1}{H}~~~,~~~\dot{R}_A=-\frac{\dot{H}}{H^2}=\frac{3}{2}(1+\frac{p}{\rho})~~~,~~~
\ddot{R}_A=\frac{9H}{2}(1+\frac{p}{\rho})(\frac{p}{\rho}-\frac{\dot{p}}{\rho})
\end{equation}

\begin{equation}
\dot{S}_A+\dot{S}_{fA}=\frac{9\pi}{2GH}(1+\frac{p}{\rho})^2
\end{equation}

\begin{equation}
\ddot{S}_A+\ddot{S}_{fA}=\frac{9\pi}{2G}(1+\frac{p}{\rho})[(1+6\frac{p}{\rho})(1+\frac{p}{\rho})-(5+3\frac{p}{\rho})\frac{\dot{p}}{\dot{\rho}}]
\end{equation}

\subsection{Event horizon:}
 
\begin{equation}
R_E=a\int_{t}^{\infty}\frac{dt}{a}~~~,~~~\dot{R}_E=HR_E-1~~~and~~~\ddot{R}_E=-\frac{4\pi G}{3}R_E(\rho +3p)-H
\end{equation}

\begin{equation}
\dot{S}_E+\dot{S}_{fE}=\frac{8\pi ^2 R_E(\rho +p)}{H}\lbrace{R_E-\frac{1}{H}\rbrace}
\end{equation}

\begin{equation}
\ddot{S}_E+\ddot{S}_{fE}=8\pi ^2(\rho +p)(R_E-\frac{1}{H})[-\lbrace \frac{R_E}{2}(1-3\frac{p}{\rho})+\frac{1}{H}+3R_E\frac{\dot{p}}{\dot{\rho}}\rbrace+R_E\lbrace1-\frac{3(1+\frac{p}{\rho})}{2(HR_E-1)}\rbrace]
\end{equation}
Note that equation (27) can be simplified further by introducing velocities at the horizons as
\begin{equation}
v_A=\dot{R}_A=\frac{3}{2}(1+\frac{p}{\rho})
\end{equation}
and
\begin{equation}
v_E=\dot{R}_E=HR_E-1
\end{equation} 
and we have
\begin{equation}
\ddot{S}_E+\ddot{S}_{fE}=8\pi ^2(\rho +p)(R_E-\frac{1}{H})[-\lbrace \frac{R_E}{2}(1-3\frac{p}{\rho})+\frac{1}{H}+3R_E\frac{\dot{p}}{\dot{\rho}}\rbrace+R_E\lbrace1-\frac{v_A}{v_E}\rbrace].
\end{equation}

We shall examine whether the above expressions for entropy variation satisfy the inequalities (12) for three DE models namely 1) Perfect fluid:~$p=\omega \rho~,~\omega <-\frac{1}{3}$, a constant, 2) Chaplygin Gas:~$p=-\frac{A}{\rho}$, $A>0$, a constant  and 3) $p=-\rho - \frac{AB\rho^{2\alpha -1}}{A\rho^{\alpha -1}+B}$, $A$, $B$, $\alpha$ are constants. It should be noted that although the third model of DE scenario is phenomenological but it contains a rich structure from the point of view of singularity. The nature of singularity for different choices of the parameters are shown below [7].\\\\ 

~~~~~~~~~~~~~~~~~~~~~~~~~~~~~~~~~~~~~{\bf Table:} Classification of singularities\\\\
\begin{tabular}{|c|c|c|}
\hline  Nature of singularity& Physical or Geometrical parameters blow up& Ranges of the parameters \\ 
\hline  Type I: The big rip& $a\rightarrow\infty , p\rightarrow\infty , \rho \rightarrow\infty$ &  $\frac{3}{4}<\alpha <1,~~A,B>0$\\ 
\hline  Type II: Sudden& $p\rightarrow\infty$& $\alpha <1~~or~~\frac{A}{B}>1$ \\ 
\hline  Type III: Big freeze& $p\rightarrow\infty , \rho\rightarrow\infty$ & $\alpha >1,~~A,B>0~or~A,B<0$ \\ 
\hline  Type IV: Big brake& $p\rightarrow\infty$ & $0<\alpha <\frac{1}{2},~~A,B>0~or~A,B<0$ \\  
\hline
\end{tabular} \\\\\\
Note that if $\alpha =0~~or~~\frac{3}{4}<\alpha <1$, then there are no type I-IV singularities. Also except type I, the other three types of singularities may occur in past or in future. Also the above four types of singularities occur when $A$ and $B$ have the same sign. However, for a difference in the sign of $A$ and $B$, there is always a sudden singularity independent of the value of $\alpha$. Further we see that if $A$ and $B$ are of same sign then $\omega <-1$ {\it i.e.,} all the four types of singularities shown above are in the phantom domain while the sudden singularity for a sign difference of $A$ and $B$ is in quintessence era. Also the second and fourth type of singularities are weak type because the geodesics can be extended beyond the singularity, the Hubble rate is finite and it occurs at finite volume. So such singularities are not of much interest from physical point of view. Hence we shall examine only type I and type III singularities in our study. Type I and type III singularities occur at finite time and at finite volume respectively and Hubble parameter and Ricci scalar diverge in both cases. As the energy density and pressure diverge in both the cases so the dominant energy condition is broken and the equation of state converges to $\omega =-1$ as universe proceeds towards the singularity.

We shall now analyze the entropy variations for the above three DE models:\\\\
\textbf{1. PERFECT FLUID:} $p=\omega \rho$, $\omega<-\frac{1}{3}$, a constant.\\

For this equation of state, equations (23),(24),(26) and (30) for entropy variations become:
\begin{equation}
\dot{S}_A+\dot{S}_{fA}=\frac{9\pi}{2GH}(1+\omega)^2
\end{equation}
\begin{equation}
\ddot{S}_A+\ddot{S}_{fA}=\frac{9\pi}{2G}(1+\omega)[(1+\omega)^2+2\omega ^2]
\end{equation}
\begin{equation}
\dot{S}_E+\dot{S}_{fE}=\frac{8\pi ^2 R_E\rho (1+\omega)}{H}\lbrace{R_E-\frac{1}{H}\rbrace}
\end{equation}
\begin{equation}
\ddot{S}_E+\ddot{S}_{fE}=8\pi ^2\rho(1+\omega)(R_E-\frac{1}{H})[-\lbrace \frac{R_E}{2}(1+3\omega)+\frac{1}{H}\rbrace+R_E\lbrace{1-\frac{v_A}{v_E}\rbrace}]
\end{equation}
Thus we have,\\\\
$~~~~~~~$ $\bullet$  $\dot{S}_A+\dot{S}_{fA}\geq0$ for all $\omega$\\

$~~$ $\bullet$  $\ddot{S}_A+\ddot{S}_{fA}>0$  in quintessence era($\omega >-1$)\\
$~~~~~~~~~~~~~~~~~~~~~~~~= 0$ in phantom barrier($\omega =-1$)\\
$~~~~~~~~~~~~~~~~~~~~~~~~<0$ in phantom era($\omega <-1$)\\\\
{\it i.e.,} the generalized second law of thermodynamics holds across the phantom barrier for universe bounded by the apparent horizon as a thermodynamical system and the dark fluid is in the perfect fluid form with constant equation of state. However, equilibrium configuration is possible only in phantom scenario, not in quintessence phase. On the other hand, for universe bounded by the event horizon as a thermodynamical system we obtain the following restrictions:\\\\

$~~~~~~~$ $\bullet$ $\dot{S}_E+\dot{S}_{fE}\geq0$ in quintessence era if $R_E>R_A$\\
$~~~~~~~~~~~~~~~~~~~~~~~~\geq0$ in phantom era if $R_E<R_A$\\
$~~~~~~~~~~~~~~~~~~~~~~~~=0$ in phantom crossing\\\\
Now,
\begin{center}
$~~~~~~~\ddot{S}_E+\ddot{S}_{fE}<0$ if GSLT holds and $R_E<\frac{2R_A}{|1+3\omega|}$ and $v_A>v_E$, {\it i.e.,}
\end{center}
In quintessence era: $R_A<R_E<\frac{2R_A}{|1+3\omega|}$, $v_A>v_E$, and\\
In phantom era:~~~~~ $R_E< min \lbrace R_A, \frac{2R_A}{|1+3\omega|}\rbrace$, $v_A>v_E$ or,\\ 
$~~~~~~~~~~~~~~~~~~~~~~~~~~~~R_E> max \lbrace R_A, \frac{2R_A}{|1+3\omega|}\rbrace$, $v_A<v_E$ and GSLT does not hold.\\\\
Thus it is possible to have an equilibrium configuration for universe bounded by the event horizon with dark perfect fluid ($\omega <-\frac{1}{3}$) both in quintessence era and in phantom era with some restrictions relating the radius of the horizons and the velocities at the horizons.\\\\
\textbf{2. CHAPLYGIN GAS:} $p=-\frac{A}{\rho}$, $A>0$, a constant.\\

In this case the sum of the entropy variation for both the horizons become:
\begin{equation}
\dot{S}_A+\dot{S}_{fA}=\frac{9\pi}{2GH}(1-\frac{A}{\rho ^2})^2
\end{equation}
\begin{equation}
\ddot{S}_A+\ddot{S}_{fA}=\frac{9\pi}{2G}(1-\frac{A}{\rho ^2})[(3\frac{A}{\rho ^2}-2)^2-3]
\end{equation}
\begin{equation}
\dot{S}_E+\dot{S}_{fE}=\frac{8\pi ^2 R_E\rho (1-\frac{A}{\rho ^2})}{H}\lbrace{R_E-\frac{1}{H}\rbrace}
\end{equation}
\begin{equation}
\ddot{S}_E+\ddot{S}_{fE}=8\pi ^2\rho(1-\frac{A}{\rho ^2})(R_E-\frac{1}{H})[-\lbrace \frac{R_E}{2}(1+3\frac{A}{\rho ^2})+\frac{1}{H}\rbrace +R_E\lbrace{1-\frac{v_A}{v_E}\rbrace}]                     
\end{equation}
Thus we have,\\\\
$\bullet$ $\dot{S}_A+\dot{S}_{fA} \geq 0$ always and\\\\
$\bullet$ $\ddot{S}_A+\ddot{S}_{fA}<0$ if $\frac{2-\sqrt{3}}{3}<{c_s}^2=\frac{A}{\rho ^2}<1$ (where $c_s$ is the velocity of sound) in quintessence era while in phantom era though the sound velocity is unrealistic ($c_{s}^2>\frac{2+\sqrt{3}}{3}$ ) but $\ddot{S}_A+\ddot{S}_{fA}<0$.\\\\
For the event horizon, we have the same restrictions for the validity of GSLT as for perfect fluid. However, for $\ddot{S}_E+\ddot{S}_{fE}<0$ we have $v_A>v_E$ and GSLT is satisfied both in quintessence era and in phantom era.\\\\
\textbf{3. $\omega=\frac{p}{\rho}$=$-1-\frac{AB\rho ^{2\alpha -2}}{A\rho ^{\alpha -1}+B}$}\\

We shall now examine the thermodynamic equilibrium configuration as the universe evolves towards a future singularity.We see that in the neighbourhood of both type of singularities ({\it i.e.,} Type I and Type III), $\dot{S}_h+\dot{S}_f\geq0$ but as $\omega \rightarrow -1$, i.e., $\dot{S}_h+\dot{S}_f\rightarrow 0$. Hence GSLT is valid in the neighbourhood of the singularities for both the horizons but as we proceed to the singularities, GSLT fails to satisfy the universe bounded by both the horizons.Also, the property of the total entropy function to have a negative second derivative is destroyed at the singularities for both the horizons. Therefore the universe departs more and more from thermodynamic equilibrium as singularity is approached.\\\\

%%%%%%%%%%%%%%%%%%%%%%%%%%%%%%%%%%%%%%%%%%%%%%%%%%%%%%%%%%%%%%%%%%%%%%%%%%%%%%%%%%%%%%%%%%%%%%%%%%%%%%%%%%%%%%%%%%%%%%%%%%%%

\section{CONCLUSIONS}

The paper deals with validity of GSLT and thermodynamical equilibrium for homogeneous and isotropic FRW model of the universe bounded by apparent or event horizon. As universe bounded by the apparent horizon is a well known Bekenstein system so we have used the Bekenstein's entropy-area relation and Hawking temperature on the apparent horizon. However, universe bounded by the event horizon is not a Bekenstein system so we have used Clausius relation to determine entropy variation on the event horizon and for the temperature we have used the recently proposed modified Hawking temperature. For the dark fluid bounded by the horizons, the entropy variation is obtained using Gibb's equation. 

The validity of GSLT on the apparent horizon for the cases under consideration is nothing new as it is commonly known irrespective of any fluid distribution. Also the restrictions for the validity of GSLT on the event horizon are already derived in [2]. However, study of thermodynamic equilibrium and its relation to GSLT and use of modified Hawking temperature on the event horizon are distinct results. In case of apparent horizon, we see that thermodynamic equilibrium is not possible in quintessence era for perfect fluid model while for Chaplygin gas model, equilibrium is possible in quintessence era for a restricted range of the sound velocity. In case of event horizon, thermodynamical equilibrium both in quintessence and phantom era is related to restrictions among the radius of the two horizons (or validity of GSLT) and the velocities at the horizons. Further, both the horizons behave in a similar fashion near the singularity and thermodynamical equilibrium is destroyed at the singularity. Thus, although universe bounded by the event horizon is not a Bekenstein system but from thermodynamical viewpoint it is not reasonable to give preference to the apparent horizon compared to event horizon, rather for perfect fluid model universal thermodynamics with event horizon has equilibrium configuration in quintessence era but it is not true for apparent horizon. Therefore, we may conclude that event horizon with the modified Hawking temperature has an edge over apparent horizon at least for the fluid models described here. For future work, appropriate definition of entropy on the event horizon may reveal more interesting phenomena.\\\\

%%%%%%%%%%%%%%%%%%%%%%%%%%%%%%%%%%%%%%%%%%%%%%%%%%%%%%%%%%%%%%%%%%%%%%%%%%%%%%%%%%%%%%%%%%%%%%%%%%%%%%%%%%%%%%%%%%%%%%%%%%%%

\begin{acknowledgments}

One of the authors(SC) is thankful to IUCAA for warm hospitality as a part of the work has been done during a visit. Also SC is thankful to DRS programme in the Department of Mathematics, Jadavpur University. The author SS is thankful to DST-PURSE Programme of Jadavpur University for awarding JRF. 

\end{acknowledgments}

%%%%%%%%%%%%%%%%%%%%%%%%%%%%%%%%%%%%%%%%%%%%%%%%%%%%%%%%%%%%%%%%%%%%%%%%%%%%%%%%%%%%%%%%%%%%%%%%%%%%%%%%%%%%%%%%%%%%%%%%%%%%


\frenchspacing
\begin{thebibliography}{7}

\bibitem{Wang1} B. Wang, Y. Gong and E. Abdalla, {\it Phys. Rev. D}  {\bf 74}, 083520 (2006).
\bibitem{Mazumder1} N. Mazumdar and S. Chakraborty, {\it  Class. Quantum. Grav.}  {\bf 26}, 195016 (2009); {\it  Gen. Relt. Grav.}  {\bf 42}, 813 (2010); {\it  Eur. Phys. J. C}  {\bf 70}, 329 (2010);   S. Chakraborty, N. Mazumder and R. Biswas,  {\it Eur. Phys. Lett.}  {\bf 91}, 4007 (2010); {\it  Gen. Relt. Grav.}  {\bf 43}, 1827 (2011); J. Dutta and S. Chakraborty,  {\it  Gen. Relt. Grav.}  {\bf 42} 1863, (2010).
\bibitem{Chakraborty} S. Chakraborty,   {\it arXiv:} 1206.1420[gr-qc]. 
\bibitem{Bennet1} C.L. Bennet etal., {\it Astron. and Astrophy.}  {\bf 399}, L19 (2003);  {\bf 399}, L25 (2003);   P. de Bernardis etal.,  {\it Nature}  {\bf 404}, 955 (2000); Gold etal.,  {\it APJs}  {\bf 192}, 15 (2011).
\bibitem{Pavon1} D. Pavon and W. Zimdahl,  {\it Phys. Lett. B} {\bf 708}, 217 (2012); H.B. Callen, {\it Thermodynamics} (J. Wiley, N.Y., 1960).
\bibitem{Izquierdo1} G. Izquierdo and D. Pavon,   {\it Phys. Lett. B}  {\bf 633}, 420 (2006).
\bibitem{Singh1} P. Singh and F. Vidotto,   {\it  Phys. Rev. D}  {\bf 83}, 064027 (2011); S. Nojiri, S.D. Odintsov and S. Tsujikawa,    {\it Phys. Rev. D}  {\bf 71}, 063004 (2005).

\end{thebibliography}
\end{document}